\documentclass{aa}
\usepackage{natbib,graphicx}
\usepackage{rotating}
\usepackage{amssymb, amsmath}
\usepackage{txfonts}
\def\mathnew{\mathsurround=0pt}
\def\simov#1#2{\lower .5pt\vbox{\baselineskip0pt \lineskip-.5pt
\ialign{$\mathnew#1\hfil##\hfil$\crcr#2\crcr\sim\crcr}}}

\def\MeV{Me\kern-0.11em V}
\def\keV{ke\kern-0.11em V}

\def\los{{\rm los}}

\begin{document}

\title{The galaxy luminosity function of the Abell~496 cluster and its
  spatial variations
\thanks{Based on observations obtained with MegaPrime/MegaCam, a joint project
of CFHT and CEA/DAPNIA, at the Canada-France-Hawaii Telescope (CFHT) which
is operated by the National Research Council (NRC) of Canada, the Institut
National des Sciences de l'Univers of the Centre National de la Recherche
Scientifique (CNRS) of France, and the University of Hawaii. This work is
 also partly based on data products produced at TERAPIX and the Canadian
Astronomy Data Centre as part of the Canada-France-Hawaii Telescope Legacy
Survey, a collaborative project of NRC and CNRS.}}

\author{G. Bou\'e \inst{1} \and
C. Adami \inst{2} \and
F. Durret \inst{1,3} \and
G. A. Mamon \inst{1,4} \and
V. Cayatte \inst{5}
}

\offprints{G. Bou\'e \email{boue@iap.fr}}

\institute{
Institut d'Astrophysique de Paris (UMR 7095: CNRS \& Universit\'e Pierre 
et Marie Curie), 98bis Bd Arago, F--75014 Paris, France
\and
LAM, Traverse du Siphon, F--13012 Marseille, France
\and
Observatoire de Paris, LERMA, 61 Av. de l'Observatoire, F--75014 Paris, France
\and
Observatoire de Paris, GEPI (UMR 8111: CNRS \& Universit\'e Denis Diderot),
61 Av. de l'Observatoire, F--75014 Paris, France 
\and
Observatoire de Paris, section Meudon, LUTH, CNRS-UMR 8102, Universit\'e Paris
7, 5 Pl. Janssen, F--92195 Meudon, France
}

\date{Accepted 29 November 2007 Received 26 April 2007; Draft printed: \today}

\authorrunning{Bou\'e et al.}

\titlerunning{Galaxy luminosity function of Abell 496}

\abstract
{The faint-end slopes of galaxy luminosity functions (LFs) in clusters
of galaxies have been observed 
in some cases
to vary with clustercentric distance
and should be influenced by physical processes (mergers, tides)
affecting cluster galaxies. However, there is a wide disagreement on the
values of the faint-end LF slopes, ranging from $-1$ to $-2.3$ in the
magnitude range $-18 < M_r < -14$.}
{We investigate the LF in the very relaxed cluster Abell~496.}
{Our analysis is based on deep images obtained at CFHT with
MegaPrime/MegaCam in four bands ($u^*$$g'$$r'$$i'$) covering a
1$\times$1 deg$^2$ region, which is centered on the cluster Abell 496
and extends to near its virial radius.  The LFs are estimated by
statistically subtracting a reference field taken as the mean of the 4
Deep fields of the CFHTLS survey.  Background contamination is
minimized by cutting out galaxies redder than the observed Red
Sequence in the $g'-i'$ versus $i'$ colour-magnitude diagram.  }
{In Abell~496, the global LFs show a faint-end slope of
  $-1.55\pm0.06$ and vary little with observing band.
Without colour cuts, the LFs are much noisier but not significantly
  steeper. 
The faint-end slopes 
show a statistically significant steepening from
 $\alpha=-1.4\pm0.1$
in the central region (extending to half a virial radius) to
$-1.8\pm0.1$ in
the Southern envelope of
the cluster.
Cosmic variance and uncertain star-galaxy separation are our main
    limiting factors in measuring the 
    faint-end of the LFs.
The large-scale environment of Abell~496, probed with the fairly
complete 6dFGS catalogue, shows a statistically significant 36 Mpc long 
filament at $\rm
PA=137^\circ$, 
but we
do not find an enhanced LF along this axis.
}
{Our LFs do not display the large number
of dwarf galaxies ($\alpha \approx -2$) inferred by 
several authors, whose analyses 
may suffer from field contamination 
caused by inexistent or 
inadequate colour cuts.
Alternatively, different clusters may have different faint-end slopes,
  but this is hard to reconcile with the 
wide range of slopes found for given clusters and for wide sets of
  clusters. 
}


\maketitle

\section{Introduction}\label{sec:intro}

Clusters of galaxies represent an extreme environment for galaxy
evolution, either in situ or through the accretion of galaxies within
groups, which are situated in the filamentary network of our
hierarchical Universe.

The analysis of the galaxy luminosity function (LF) in several
wavebands is a good way to sample the history of the faint galaxy
population (e.g. \citealp{Adami+07}) including star formation history,
evolutionary processes and environmental effects.  In particular, the
slope of the faint-end of the LF is a direct indicator of the importance of
dwarf galaxies, which are expected to be more fragile in the
environment of clusters.

The great majority of studies of the LF indicate faint-end slopes in
the range $-0.9$ to $-1.5$, but these mostly did not reach very faint
magnitudes (see Table~1 in \citealp{DePropris+03} and references
therein).  Recent deep imaging has shed more light on the LF at faint
luminosities.
Table~\ref{alphalit} shows deep (fainter than absolute magnitude
$M=-16$) estimations of the faint-end slope: most studies conclude to
fairly shallow slopes $\alpha \simeq -1.3$ (with typical uncertainties
  of 0.1  to 0.2), while several point to
faint-end slopes as steep as $\alpha \approx -2.3$, which diverges in
luminosity (unless the LF becomes shallower or is cut off at some very
faint luminosity).

Part of the wide range
of faint-end slopes may be caused by cosmic variance of the background
  counts. 
The range of faint-end slopes of the LF may also be due to different mass
buildup histories of clusters, through spherical and filamentary
infall and major cluster-cluster mergers.  
However, some of this
dispersion in slopes could be caused by 
systematic uncertainties such as in the
star/galaxy separation or through different surface brightness cuts.

Spectroscopic-based LFs would alleviate this problem, but until recently,
such spectroscopic-based LFs do not extend fainter than
$M=-16$. The exceptions are studies by \cite{RG08} and \cite{MMB08} who both
find shallow slopes for Virgo cluster galaxies down to $M_r = -14$, as well as
\cite{PC07} for the Perseus cluster core.

Unfortunately, galaxy formation simulations do not yet probe the galaxy LF
down to sufficiently faint luminosities: the deepest study, by
\cite{Lanzoni+05}, probed the LF with the GALICS semi-analytical galaxy
formation model only down to $M_B < -16.5$ (they found $\alpha \simeq
-1.3$ in the $B$ band and 
$\simeq -1.4$ in the $K$ band within the virial radii of clusters, and
$\alpha \simeq -1.0$ between 1 and typically 3 virial radii).

The present paper aims at clarifying the debate on the faint-end of the LF
and on
 understanding the different effects of
spherical and filamentary infall on the LF.
Indeed, the influence of infall on the
galaxy population, in particular on the LF is not always well
understood, except for a few clusters, such as Coma
(e.g. \citealp{Adami+07}).
To achieve this aim, we chose to analyse the very relaxed
cluster, Abell~496, known to be very regular both at X-ray and optical
wavelengths, and also from a dynamical point of view
\citep{Durret+00}. Our data cover a field of view that is wide enough
to reach the virial radius and thus
probe a variety of environments.  
The LFs are computed after a better filtering of artefacts through a
  minimum 
galaxy width, a better filtering of background galaxies through the
rejection of galaxies redder than the Red Sequence of early-type galaxies,
and make use of a thorough analysis of the uncertainties due to cosmic
variance, photometric errors and imperfect star/galaxy separation. 
Our general motivation is to 
obtain LFs in
various regions of Abell~496 
and compare them with previous works.

The paper is organised as follows. We present our MegaCam data and
data reduction in Section 2. In Section 3, we describe how we compute
LFs using large comparison fields from the CFHTLS to statistically
subtract the fore- and background galaxy population. In Section 4, we
present our results obtained for the LFs of Abell~496 in various
regions. In Section 5, we briefly discuss our results concerning the
LFs in 
terms of large scale environmental effects on the cluster galaxy
populations.  Finally, in Section 6 we compare our LFs to other
  determinations of the Abell~496 LF, and we discuss the discrepancy between
  our moderate faint-end slope and the steep faint-end slopes recently found
  by several authors in several clusters.

With a mean heliocentric velocity of $9885 \, \rm km \, s^{-1}$
\citep{Durret+00}, 
Abell~496 has a (luminosity) distance modulus of 35.73,
and the scale is 0.636 kpc
arcsec$^{-1}$ (including cosmological corrections\footnote{We used
the cosmological corrections in the CMB frame as  provided in the NASA/IPAC
  Extragalactic Database (NED).}).
We give magnitudes in the AB system.

\section{Observations and data reduction}

\subsection{MegaCam cluster data}
\label{data}
Abell~496 is centered at J2000 equatorial coordinates 
$04^h33^m37.1^s, -13^\circ14'46''$. 
It has an angular virial radius of
$0.77^\circ$ (virial 
radius of 1.9 Mpc), obtained by extrapolating the radius of overdensity 500
(\citealp{MVFS99}, measured relative to the critical density of the Universe)
  to the radius of 
overdensity 100.

The MegaCam field covers an area corresponding to
2.3$\times$2.3 Mpc$^2$ at the cluster redshift. 
We centered our images of Abell~496 on the cluster centre (i.e. on the
cD galaxy). 
This means that we can
cover the whole Abell~496 area and its immediate infalling layers
within the virial radius.

Abell 496 was observed at CFHT with the large field MegaPrime/MegaCam
camera in November 2003 on program 03BF12, P.I. Cayatte (see
Table~\ref{tab:observation}).  Images were reduced by the TERAPIX
pipeline using the standard reduction tool configuration. We refer the
reader to http://terapix.iap.fr/ for reduction details.

\begin{table*}
\begin{center}
\caption{Observation characteristics
\label{tab:observation}
}
\tabcolsep 4pt
\begin{tabular}{lcrrrrcllllc}
\hline
Name & 
Useful area & \multicolumn{4}{c}{Exp. time} & & \multicolumn{4}{c}{PSF} &
Obs. \\
& (deg$^2$) & \multicolumn{4}{c}{(sec)} & & \multicolumn{4}{c}{(FWHM in
  arcsec)} & Date \\ 
\cline{3-6}
\cline{8-11}
 & & $u^*$ & $g'$ & $r'$ & $i'$ &
& $u^*$ & $g'$ & $r'$ & $i'$ 
& \\
\hline\hline
Abell 496 & 0.82 & $13\,680$ & $7\,820$ & $3\,780$ & $3\,570$ &
& 1.13 & 1.06 & 0.88 & 0.94 & \ \ 11/2003 \\
Deep 1 & 0.77 & 38\,946 & 24\,893 & 60\,854 & 134\,863 & 
& 1.06 & 0.96 & 0.92 & 0.91 & $<$09/2006 \\
Deep 2 & 0.79 & 5\,281 & 16\,655 & 31\,988 & 72\,734 & 
& 0.89 & 0.96 & 0.90 & 0.91 & $<$09/2006 \\
Deep 3 & 0.83 & 19\,146 & 21\,392 & 59\,574 & 109\,049 & 
& 1.15 & 0.96 & 0.94 & 0.89 & $<$09/2006 \\
Deep 4 & 0.82 & 50\,867 & 24\,262 & 72\,736 & 140\,981 & 
& 1.03 & 0.98 & 0.88 & 0.87 & $<$09/2006 \\
W1 & ----- & 2\,215 & 2\,436 & 1\,179 & 4\,189 & 
& $1.04\pm0.09$ & $0.96\pm0.05$ & $0.86\pm0.11$ & $0.85\pm0.12$ & $<$09/2006 \\ 
\hline
\end{tabular}

Note: The W1 values are means over the 19 Wide subfields  of the Wide
  field.
\end{center}
\end{table*}

Object extraction was made using the SExtractor package \citep{BA96}
in the double-image mode.  The CFHTLS pipeline at the TERAPIX data
centre creates a $\chi^2$ image based upon the quadratic sum of the
images in the different wavebands.  Objects are then detected on this
image.  In contrast with the CFHTLS images, our set of
$u^*$$g'$$r'$$i'$ images for each of the two clusters presents
important differences in their PSFs (see Table~\ref{tab:observation}).
For this reason, we chose a different approach from that of the
TERAPIX data centre: instead of using the $\chi ^2$ image as the
reference image, we use the band with the best seeing in our data:
$r'$.  Detections were performed in this band and object
characteristics were measured in all bands. The detections and
measures were made using the CFHTLS parameters, among which an
absolute detection threshold of 0.4 ADU
above the background 
($\mu < 27.34$ in all bands), a minimal detection area of
3 pixels and a 7$\times$7 pixel gaussian convolution filter of 3
pixels of FWHM. In each of the $u^*$, $g'$, $r'$ and $i'$
output catalogues, we only kept
objects with 
semi-minor axes larger than 1 pixel and mean surface brightness within
the half-light radius greater than $\mu=26.25$ in order to remove
artefacts.

We measured Kron magnitudes ({\tt
    MAG\_AUTO} in SExtractor), 
with the default SExtractor settings.
We used the
  photometric calibration given by the TERAPIX data processing centre.
Since the fluxes in different bands are measured within the same Kron
  elliptical aperture, we derive colours by simply subtracting the
  magnitudes. Therefore, our colours are not affected by
  aperture effects and are only slightly affected
by the differences in the PSF between the two bands involved.

Using simulations, we also estimate the completeness levels and
reliabilities of our detections (see e.g. \citealp{Driver+98}).  This
is crucial, since we intend to compare our data with CFHTLS data that
were not observed exactly in the same conditions. For this, we used
the SkyMaker package \citep{BF07} to build images with the same noise
and point spread function as our MegaCam images on the one hand and
CFHTLS Deep images on the other hand. The objects fed into SkyMaker
were either spherical to flattened S\'ersic bulges\footnote{We
modified SkyMaker to handle S\'ersic profiles rather than just the de
Vaucouleurs profile.} or thin exponential disks.  The S\'ersic shape
and effective radius are specified functions of luminosity that
\cite{ML05} obtained from the luminous galaxies of Abell~496 and Coma
data of \cite{Marquez+00}, while for the dwarfs we considered the
relations given by \cite{BJ98}. 
The central $B$-band 
disk surface magnitudes
were extrapolated from a gaussian distribution
$\mu_B(0) = 21.5\pm1$.
The
fraction of ellipticals is 60\% in clusters and null in the field.
The galaxy luminosity function was taken from \cite{PBBR06} for the
cluster and from \cite{Blanton+03} for the field.

Completeness (ratio of real detections to real sources) and
reliability (ratio of real to total detections) were then measured by
cross-correlation between the SkyMaker input and the SExtractor output
catalogue. 
We found that up to $i'$=23, for both kinds of images,
SExtractor finds 80\% of the objects, and among the detected objects,
only 10\% are artefacts.

\begin{figure}[ht]
\centering
\includegraphics[width=8cm,angle=270]{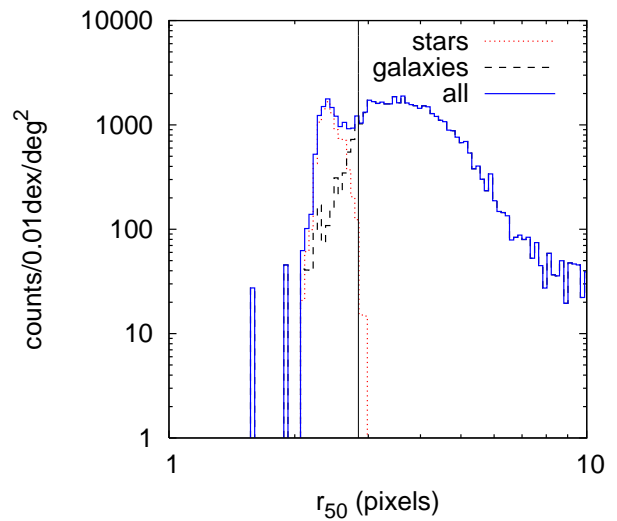}
\caption[]{Distribution of half-light radii in the magnitude bin $22 < r' <
  22.5$. The \emph{vertical line} shows the star/galaxy separation deduced
  from the Besan\c{c}on model.
\label{fig:bimodal}}
\end{figure}

We then performed a star-galaxy separation.  Instead of using the
neural-network star-galaxy classification method of SExtractor, we
placed the detections in a diagram of size (i.e. the half light
radius) versus magnitude in the band with the best seeing.
The variation of the PSF over the images has
been corrected for using a technique similar to that of
\cite{McCracken+03}.  Stars are the smallest objects and are located
in a well defined strip up to $r'$$\approx$21, thus allowing the
separation. At fainter magnitudes, stars and galaxies overlap and
individual classification is no longer possible. 
For magnitudes range where the half light radius distribution is
  bimodal, we perform a statistical star/galaxy separation,
assuming gaussian distributions of $\log
r_{50}$ for both stars and galaxies, in different magnitude bins.
This is illustrated in Fig.~\ref{fig:bimodal}
for the fairly faint magnitude bin $21 < r' < 21.5$.

\begin{figure}
\centering
\includegraphics[width=8cm,angle=270]{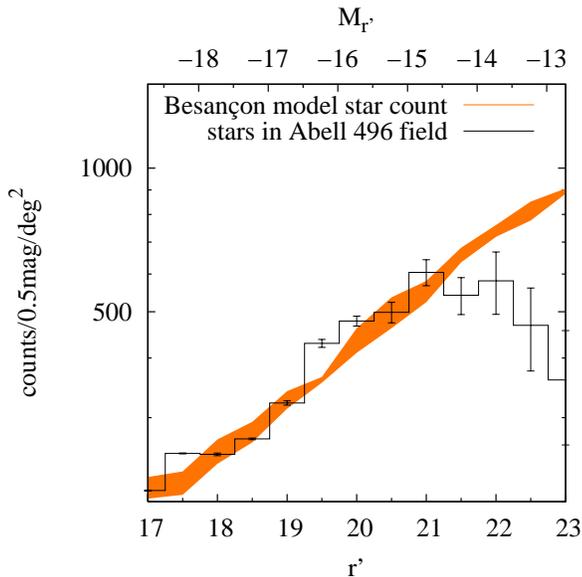}
\caption[]{
Comparison
of star counts obtained with the Besan\c{c}on model \citep{RRDP03}
(\emph{orange shaded region}) and from the Abell~496 image 
(\emph{black solid histogram}) using our star-galaxy separation. 
}
\label{fig:besancon}
\end{figure}

Fig.~\ref{fig:besancon}  
compares our star counts with those from 
the Besan\c{c}on model (\citealp{RRDP03}). We
ran the Besan\c{c}on model six times in order to get the error on these 
counts. We did the same with our algorithm. At bright magnitudes ($r' <
  21$), the two distributions coincide.
However, at faint magnitudes ($r'>21$), our star counts decrease, while
the 
Besan\c{c}on model counts keep increasing. 
This could be due to a steeper fall of the stellar density
  distribution in the direction of Abell~496, in comparison with what is in
  the 
Besan\c{c}on model. 
Alternatively, we may be wrongly classifying stars as galaxies at
  $r' > 21$, and at $r' = 22$ 
we may be overestimating the galaxy counts in
  the Abell~496 field. If we adopt the star counts from the Besan\c{c}on
  model, we would end up with 114, 156 and 343 fewer galaxies at  $r'$
  = 
  21.5, 22.0 and 22.5, respectively.
However, if there were as many stars as predicted by the Besan\c{c}on
  model, then, for $22 < r' < 22.5$ (see Fig.~\ref{fig:bimodal}), the
  distribution of stellar half 
  light radii would rise to a maximum (near $r_{50}=1.3$ pixels) fall to a
  minimum (near 1.6 pixels), then rise again (to the limit of 1.8
  pixels). It is difficult to understand what would cause this final rise.
Therefore,
in what follows, we adopt our own estimation of the star counts.

We also corrected magnitudes for Galactic extinction based on the
\cite*{SFD98} maps.  We finally computed the useful covered area (cf.
Table~\ref{tab:observation}), by masking all saturated stars,
spikes 
and CCD edges.
Galaxies and stars brighter than $r'=18$
occupy less than 2\% of the pixels of our image.

The final catalogue will be electronically available at the following
address: http://www.cencos.fr/ and in a few months the images will be
available at the same address.

\subsection{CFHTLS comparison field data}

We used the CFHTLS Deep (D1, D2, D3 and D4, i.e. 4 MegaCam fields) and
Wide (W1, W2 and W3, 59 MegaCam fields) as comparison field data.
Because the LFs are computed by subtraction of the average of the 4
Deep fields (DFs) from the Abell~496 field, \emph{objects were
re-extracted from the 4 DFs in exactly the same manner as in the
cluster field}: i.e., with the same detection waveband ($r'$ rather than
$\chi^2$ images) the
same SExtractor parameters 
(including the same 0.4 ADU threshold, which given
the same zero point corresponds to the same absolute threshold), and the 1
pixel minimum semi-minor axis. 
Since the Deep images are deeper than the Abell~496 images, we apply the
  same cut of mean surface magnitude within the half-light radius of
$\mu<26.25$ for all 4 wavebands.

We chose the DFs as reference fields (rather than the CFHTLS Wide
fields) because their greater depth ensures smaller photometric errors
in our range of magnitudes than those of the Wide fields.  Moreover,
the DFs were selected to be free of rich nearby structures, which is
not the case for the Wide CFHTLS fields, which are shallower (except in
  $i'$) than the Abell~496 field. 

\begin{figure}[ht]
\centering
\includegraphics[width=8cm,angle=270]{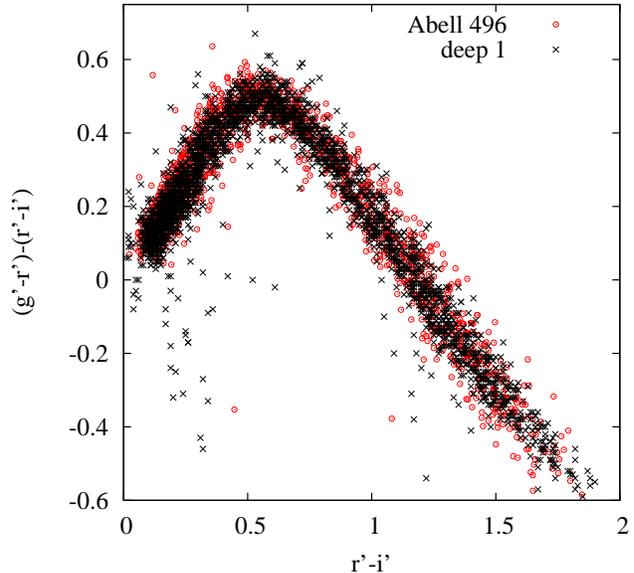}
\caption{Colour-colour diagram for $17 < r' < 20$ stars in the A496
  (\emph{open red circles}) and D1 (\emph{black crosses}) fields, corrected for
  extinction, and with a shift of the D1 $i'$ photometry by +0.02
  magnitude. 
\label{fig:colcol}}
\end{figure}

We checked that the cluster and reference fields have compatible
  photometric calibration. For this we made a colour-colour diagram, shown in
  Fig.~\ref{fig:colcol}, in which we corrected the magnitudes for galactic
  extinction using the \cite{SFD98} model, and shifted the D1 $i'$
  magnitudes by +0.02 to force a match.

\begin{figure}
\centering
\includegraphics[width=7.3cm,angle=270]{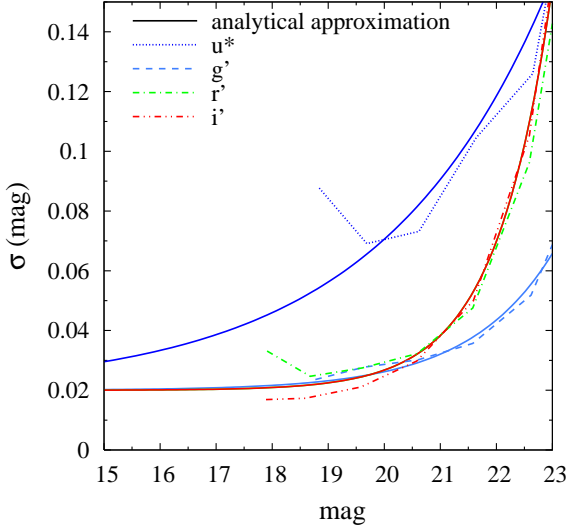}
\caption[]{Magnitude uncertainties estimated from overlapping areas in the
19 CFHTLS W1 fields \emph{(dotted and dashed curves)} with their analytical 
approximation (eq.~[\ref{sigmfit}], \emph{solid curves} ).}
\label{fig:err}
\end{figure}

The W1 region (19 MegaCam fields) of the CFHTLS was
considered to estimate the magnitude 
uncertainties as a function of magnitude in an external way. For this,
we considered the overlapping areas of the 19 W1 fields. In these
areas, we compiled the objects observed twice, and this allowed to
estimate the magnitude difference as a function of magnitude in the
$u^*$, $g'$, $r'$ and $i'$ bands. We only selected objects located
more than 400 pixels away from the field edges in order to avoid
artificially increasing the magnitude uncertainties due to border
effects. 
These uncertainties are shown in Fig.~\ref{fig:err}.

We used the TERAPIX object catalogues for the W1 fields, for which detections
  were done on the $\chi^2$ images. In each waveband, the W1 fields
have the same measurement threshold as the Abell 496 image. However,
  the measurement
isophote is noisier in the W1 fields (except in the $i'$ 
band).
but with the same area on
average. Therefore, the W1 magnitude uncertainties should be upper
  limits for the Abell~496 magnitude uncertainties.

These uncertainties can be approximated by
\begin{equation}
\sigma_m=0.02+\alpha\exp\left[\frac{m-m_1}{\delta_m}\right] \ ,
\label{sigmfit}
\end{equation}
where
$(\alpha, m_1, \delta_m)$ = (1.2, 29.5, 3.0), (2.5, 29.0, 1.5),
(1.0, 25.0, 1.0) and (1.0, 25.0, 1.0) for $u^* g' r' i'$ respectively.
These expressions are used to compute the uncertainties on galaxy counts.
Note that although the W1 pointings had comparable PSFs to the Abell~496
  PSFs, the integration times were smaller, except in the $i'$ band (see
  Table~\ref{tab:observation}). Hence, the photometric errors derived
    from the W1 field are
  upper limits in the $u^*$, $g'$ and $r'$ bands.

We also recomputed the area coverage for the Deep and Wide fields
from the CFHTLS mask files
(see Table~\ref{tab:observation}).

\section{Description of the methods}

\subsection{Luminosity function calculations}

The basic method is to statistically estimate the fore- and background
contributions to the cluster lines of sight using comparison fields
free of rich nearby structures
\citep{Oemler74}.

\begin{figure}
\centering
\includegraphics[width=8cm,angle=270]{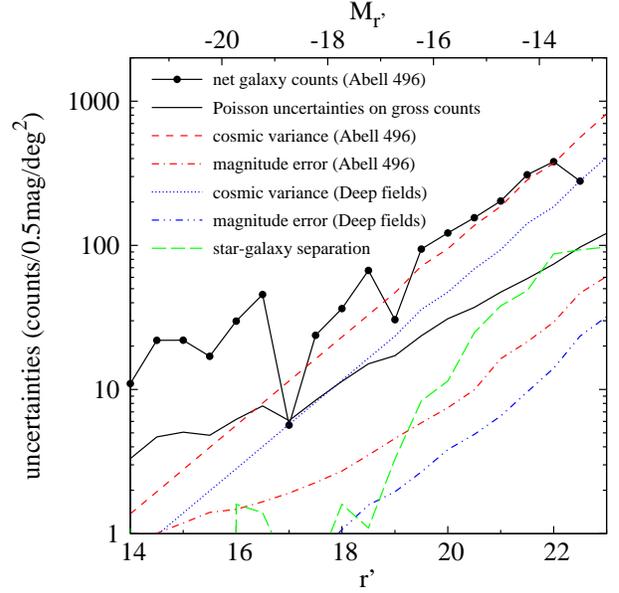}
\caption[]{Uncertainties on the global LF of Abell 496 in the $r'$
band for all galaxies, i.e. without selection from the colour-magnitude
diagram. For comparison, we show the Poisson error calculated on the
gross counts of Abell 496. Note that we defined cosmic variance
  (eq.~[\ref{sigcv}]) with the
Poisson contribution.
}
\label{fig:allsig}
\end{figure}

We compute the LF in the standard fashion: we subtract the reference field
counts (the mean of the 4 DFs) from the cluster field counts.
The uncertainty is estimated as follows.
In a first step, we compute the uncertainties coming from errors on
magnitude measurements: starting from a catalogue of magnitudes $\{m_i\}$, we
create mock catalogues $\{m^\prime_i\}$, where
$m^\prime_i$ are gaussian distributed random variables of mean
$m_i$ and standard deviation $\sigma(m_i)$ estimated from 
overlapping areas in the 19 CFHTLS W1 fields (cf. Fig.~\ref{fig:err}).
Uncertainties on distributions are then
$\sigma_m=\sigma\{N(m^\prime)\}$.
In a second step, we compute the uncertainties due to the cosmic
variance using the 59 CFHTLS Wide fields: 
\begin{equation}
\sigma_{\rm CV}^2(m) = \sigma^2\{N_{\rm Wide}(m)\} - 
\langle \sigma_m^2(m) \rangle_{\rm Wide} \ ,
\label{sigcv}
\end{equation}
where $\langle\rangle_{\rm Wide}$ means the median over the 59
Wide fields. It should be stressed that $\sigma_{\rm CV}(m)$ defined in
this way implicitly takes into account the statistical
uncertainties on the 
star-galaxy separation (cf. Fig.~\ref{fig:besancon}).
Finally, we add quadratically all uncertainties: 
\begin{eqnarray}
\sigma^2\{N^{\rm Cl}(m)\} &=& 
\sigma_m^2\{N^{\rm Cl}_\los(m)\} + 
\sigma_{\rm CV}^2(m) \nonumber \\
&&+\,\frac{1}{4}\left[\sigma_m^2\{N^{\rm DF}(m)\}
+ \sigma_{\rm CV}^2(m)
\right] \ ,
\label{sigcts}
\end{eqnarray}
where $N^{\rm Cl}$, $N^{\rm Cl}_{\rm los}$ and $N^{\rm DF}$ are the counts
from the cluster, the cluster field and the mean of the 4 DFs, respectively.
The factor of 4 corresponds to the 4 DFs. All these uncertainties are 
plotted in Fig.~\ref{fig:allsig} for Abell~496 in the $r'$ band.

The uncertainty in the galaxy counts given in equation~(\ref{sigcts})
  as well as the contribution of star/galaxy separation to the uncertainty in
  the   counts   shown
in Fig.~\ref{fig:allsig} 
are
  purely statistical. One should also consider systematic
  contributions to this uncertainty, in particular those from  star/galaxy
  separation. Indeed, the difference in galaxy counts after subtraction of
  the stars either from our star counts or from the Besan\c{c}on model is
  almost as large as the cosmic variance of the galaxy counts of the
  Abell~496 cluster field. 
However, our analysis of Sect.~\ref{data} suggests that this
  systematic uncertainty is smaller than the difference between our estimated
  star counts and those obtained with the Besan\c{c}on model.

\subsection{Improvement using colour magnitude relations}

LFs computed directly from the method described above show very big
error bars mainly due to the cosmic variance (cf. Fig.~\ref{fig:allsig}). 
This problem has already been highlighted
(e.g. \citealp{Oemler74,DAL02}). We improve our analysis by
  removing those objects whose colour
imply that they are background objects.
We considered $g'-i'$ because it corresponds to the highest quality
magnitude wavebands.

\begin{figure}
\centering
\includegraphics[width=8cm,angle=270]{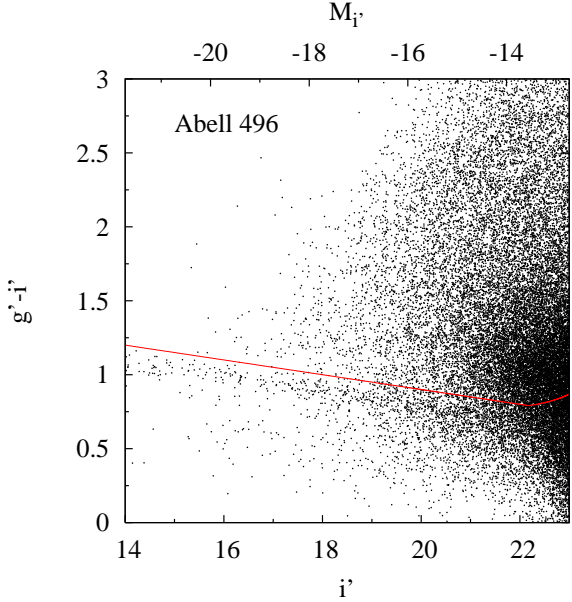}
\caption{$i'/g'-i'$ colour-magnitude diagram for Abell 496. The \emph{red curve}
shows the upper limit of galaxy selection to compute the LFs.
}
\label{fig:cmr}
\end{figure}

Fig.~\ref{fig:cmr} shows the colour magnitude relations of Abell 496,
where no background subtraction has been made.  We see a well defined \emph{Red
Sequence} that decreases linearly with $i'$, down to at least $M_i =
-14.5$, as
\begin{equation}
(g'-i')_{\rm RS} \simeq 1.75 - 0.05\,i' \ ,
\label{redseq}
\end{equation}
 consistent with what has been
known since \cite{BLE92}. We assume that this Red Sequence is real and
corresponds to the reddest galaxies of the cluster.  We therefore
select only galaxies slightly (0.15 magnitude) above our Red Sequence.
Because our photometric errors increase with magnitude, the strict
application of this colour cut would lead to an incompleteness at the
faint-end. Therefore we use the cut
\begin{eqnarray}
g'\!-\!i' &\!\leq\!& (g'\!-\!i')_{\rm RS} + \hbox{Max} \left
(0.15,1.5\,\sigma_{g'\!-\!i'} \right 
) \nonumber \\
&\!=\!& 1.75-0.05\,i' 
\nonumber \\
&\mbox{}& \qquad + \hbox{Max} \!
\left [0.15, 1.5\,\sqrt{\sigma_g^2 \left (0.95\,i\!+\!1.75 \right) + \sigma_i^2(i)}
    \right ] \ ,
\label{coloureq}
\end{eqnarray}
where we wrote $\sigma^2_{g'-i'} = \sigma_{g'}^2 + \sigma_{i'}^2$, 
used the Red Sequence (eq.~[\ref{redseq}])  to translate $g'$ to $i'$, and
took $\sigma_g(g')$ and $\sigma_i(i')$ from equation~(\ref{sigmfit}).
The factor 1.5 in the first equality of equation~(\ref{coloureq}) ensures that
we are 93\% complete (assuming a gaussian probability distribution function,
hereafter pdf).

The colour cut of equation~(\ref{coloureq})  is represented in
Fig.~\ref{fig:cmr} by the red curve. 
While our colour cut may lead to a loss of atypically red cluster
galaxies (e.g. dusty objects), we are confident that such a
population, if it exists, is small, and will only marginally decrease
our completeness.  On the other hand, the colour cut will drastically
improve our reliability in the net cluster counts (indeed, our tests have
shown that without the colour cuts, the LF is much noisier).  Hereafter, all
LFs as well as the cosmic variance are 
computed using this selection.

\section{Results}

We computed LFs both for the whole field of view and for
16 subfields. The subfields define a regular square grid of
15$\times$15~arcmin$^2$ each and allow a good compromise between
spatial resolution and uncertainties in individual magnitude bins. We
used 1~magnitude bins to limit the uncertainties. Several subregions
are then defined including a certain number of subfields with common
properties; they were chosen without assuming circular symmetry for the
cluster.

\begin{figure}
\centering
\includegraphics[width=10cm,angle=270]{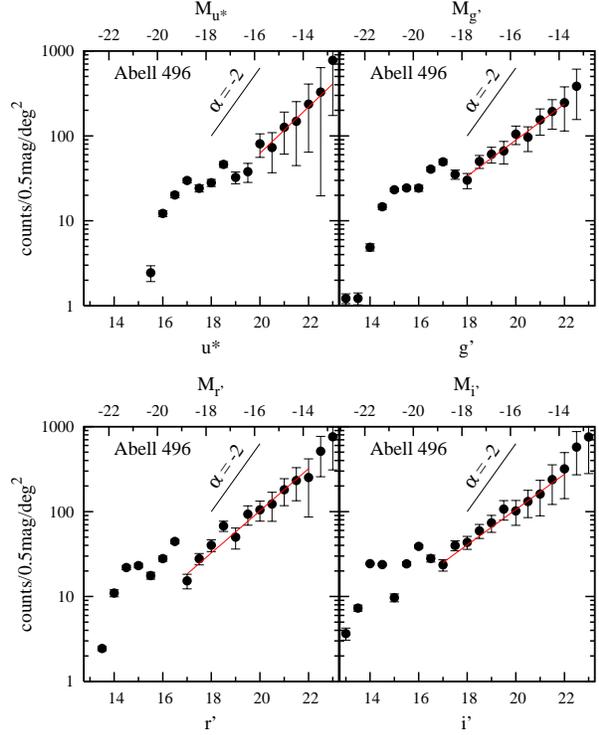}
\caption[]{Global luminosity functions for Abell 496 in the four bands
with the best fits (in red).}
\label{fig:glf496}
\end{figure}

\begin{table*}
\begin{center}
\caption{Faint-end slopes of Abell~496}
\begin{tabular}{l*4{c}}
\hline
\hline
Region & $20  \leq u^*\leq23  $ &
  $18  \leq g'\leq22  $ & 
  $17  \leq r'\leq22  $ &
  $17  \leq i'\leq22  $ \\
& $-15.73  \leq M_{u^*}\leq -12.73  $ &
  $-17.73  \leq M_{g'}\leq -13.73  $ & 
  $-18.73  \leq M_{r'}\leq -13.73  $ &
  $-18.73  \leq M_{i'}\leq -13.73  $ \\
\hline
All &
    ${-1.68\pm0.35}$ &
    ${-1.53\pm0.08}$ &
    ${-1.62\pm0.05}$ &
    ${-1.52\pm0.05}$ \\ 
All (no colour cuts) &
    ${-1.78\pm0.51}$ &
    ${-1.61\pm0.17}$ &
    ${-1.73\pm0.18}$ &
    ${-1.58\pm0.24}$ \\ \hline
Centre &
    ${-1.60\pm0.27}$ &
    ${-1.43\pm0.07}$ &
    ${-1.39\pm0.08}$ &
    ${-1.41\pm0.05}$ 
\\ 
South &
    ${-1.87\pm0.34}$ &
    ${-1.89\pm0.14}$ &
    ${-1.79\pm0.11}$ &
    ${-1.80\pm0.10}$ 
\\ 
East-North-West &
    ${-1.88\pm0.69}$ &
    ${-1.48\pm0.22}$ &
    ${-1.40\pm0.25}$ &
    ${-1.58\pm0.12}$ \\ 
\hline
\hline
\end{tabular}
\end{center}

Note: the magnitude intervals correspond to the \emph{centres} of 0.5 magnitude
bins (All) and 1.0 magnitude bins (Centre, South, East-North-West).
\label{tab:LF}
\end{table*}

The LFs of Abell~496 in the four bands are displayed in
Fig.~\ref{fig:glf496} (the corresponding data are given in Table~.2 of
the appendix), 
showing that the shapes of the global LFs of the Abell~496 field are
similar in the four bands, with the faint-ends increasing linearly.  As
these LFs do not look like Schechter functions, we decided 
to fit only the faint-ends by a power-law. The expression in terms of
magnitude is given by:
$$ \Phi(M) = 10^{-0.4(\alpha+1)(M-M_0)}. $$ 
The best fit slopes $\alpha$ of the overall LFs are given in
Table~\ref{tab:LF}. We used the Levenberg-Marquardt method
(e.g. \citealp{NumRecC2})
to fit the
data. Error bars were computed from 1000 parametric bootstraps,
where the pdf of the net counts is
assumed gaussian with a width obtained from the pdf of the gross galaxy 
counts
of the 59 Wide
fields.
The faint-end slopes vary from band to band, but are typically
$\alpha=-1.55\pm0.05$. 

We recomputed the LF in the $r'$ band 
using a much more conservative cut in
  surface brightness: $\mu_r < 24.25$ (instead of 26.25). The faint-end slope
  becomes $\alpha = -1.60\pm0.05$ (instead of $-1.62\pm0.05$).
Hence, the faint-end slopes appear robust to different cuts in surface
  brightness.

Had we adopted instead the star counts from the Besan\c{c}on model (see
  Fig.~\ref{fig:besancon}), the slope of the LF in the $i'$ band for the
  global field would have been $-1.27\pm0.04$.

We also computed the LFs for galaxies in the Red Sequence (where the redder
limit is taken from eq.~[\ref{coloureq}], while the bluer limit is the
symmetrical cut, with respective to the average Red Sequence given in
eq.~[\ref{redseq}]). The slopes for the Red Sequence galaxies matched those of
the global LFs in all bands. The LFs for galaxies bluer than the Red 
Sequence are not significantly different but with larger error bars 
($\alpha\approx -1.65\pm0.15$).

\begin{figure}
\centering
\includegraphics[width=8cm,angle=270]{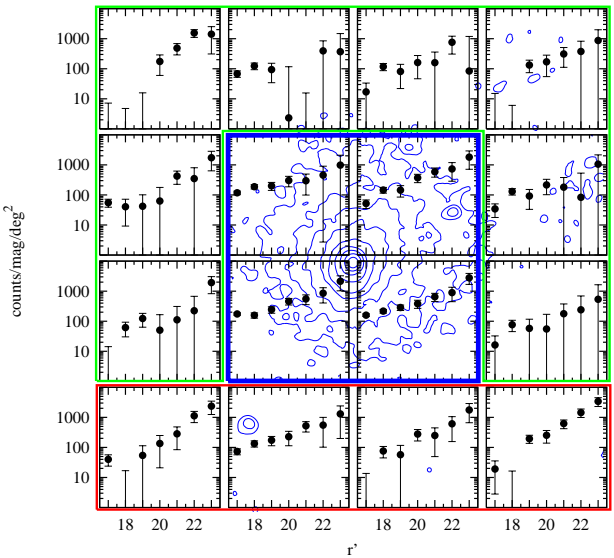}
\caption[]{Local luminosity functions for Abell 496 in the $r'$
band. Each subfield is 15$\times$15~arcmin$^2$. X-ray intensity
contours with logarithmic steps from ROSAT/PSPC data are superimposed.
Three main areas are defined in \emph{blue}, \emph{red} and
\emph{green}, which correspond respectively to the centre, the well
populated Southern rectangle and the sparse Northern ring.}
\label{fig:llf496}
\end{figure}

The local $r'$ LFs in the 16 subfields of Abell~496 are displayed in
Fig.~\ref{fig:llf496}. This figure shows that the LFs are not similar
over the whole cluster field: subfields in the North, East and West
extremities of the cluster are sometimes poorly populated and exhibit
large error bars, while subfields in the Southern region show rising
LFs.  We can thus divide the cluster into three main regions: a
central region 30$\times$30 arcmin$^2$ (1.15$\times$1.15 Mpc$^2$, in
blue in Fig.~\ref{fig:llf496}), an East-North-West region around this
central zone (in green in Fig.~\ref{fig:llf496}) and a Southern region
(in red in Fig.~\ref{fig:llf496}).  The central region extends to
roughly one half of the virial radius
and corresponds to the densest region of the cluster.  

\begin{figure}
\centering
\includegraphics[width=10cm,angle=270]{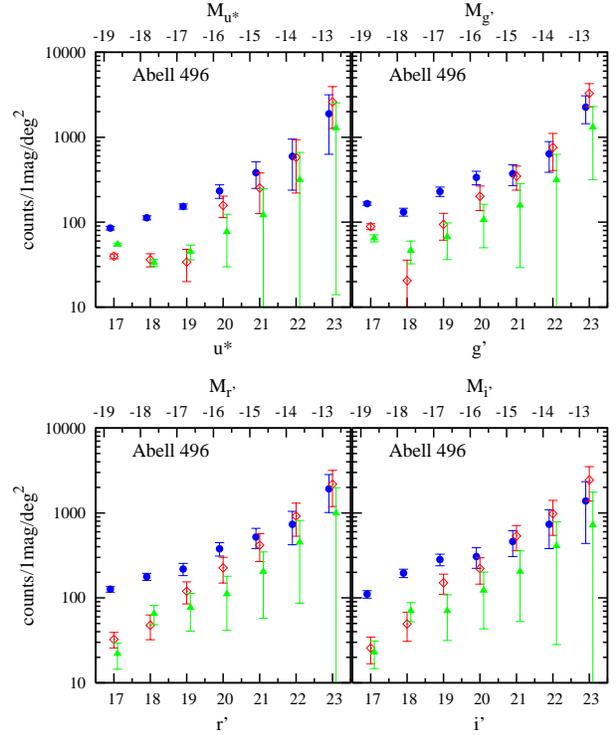}
\caption[]{Luminosity functions for Abell 496 in the four bands and 
in the three main areas defined Fig.~\ref{fig:llf496}. The blue colour
corresponds to the central region, red is for the South and green
for the upper East-North-West zone.
}
\label{fig:lglf496}
\end{figure}

Fig.~\ref{fig:lglf496} shows the LFs computed 
in these three subregions, in the four photometric
bands (the corresponding data are given in Table~.2 of 
the appendix). 
The LFs have faint-end slopes (see Table~\ref{tab:LF}) that
are significantly
shallower in the centre than in the Southern periphery in all bands except $u^*$.
The Southern region has a surface density of faint galaxies
  ($M_{r'} > -14$) that is higher than or comparable to that of the central
  region. 
If this is not caused by field
  contamination (see Sect.~\ref{ubiq}), then one would conclude that the
  faint ($M_{r'} > -14$) galaxies do not trace the cluster, which presents no
  surface density enhancement.

Although they are both contained within the virial radius of the cluster,
the two
external regions present LFs differing from one another. The Southern
region (red) is still quite populated  compared to the green region,
which is sometimes quite poor with LFs often not significantly
positive. 
Since there are no clusters or groups known nearby (as searched with
  NED, see Fig.~\ref{fig:lss}),
this suggests there may be matter in the Southern region
infalling from the surrounding cosmological web, as discussed in the
next Section.

\section{Large scale filament in the neighbourhood Abell~496}

\begin{figure}
\centering
\includegraphics[width=\hsize]{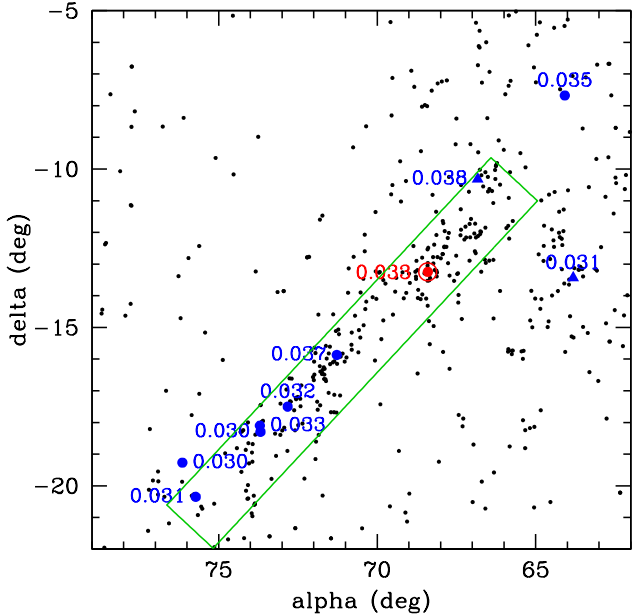}
\caption{Large scale structure surrounding Abell 496,
in a region of 17$\times$17~deg$^2$ (41$\times$41 Mpc$^2$), in a redshift slice
 within $\pm 0.005$ of that of the cluster.
The galaxies  are taken from  6dFGS-DR3
  \emph{(points)}, limited to the completeness limit of $K_s < 12.65$, which
  corresponds to $L^*/4$ (using the $K_s$ field LF of \citealp{JPCS06}).
We also show groups (\emph{blue triangles}) and clusters 
(\emph{large blue circles})
found in 
NED in the same redshift interval (with their redshifts highlighted).
The \emph{surrounded red circle} shows the position of the cluster centre.
}
\label{fig:lss}
\end{figure}

Abell~496 has been shown to be a noticeably
quiescent and relaxed cluster (e.g. \citealp{Durret+00}). 
The only sign of substructure found was
an enhanced concentration of emission line galaxies in the
North-West. From Fig.~\ref{fig:llf496}, we also see that the Southern part
of the surrounding area around the cluster shows a 
significant galaxy
population, with an LF that resembles that of the central cluster.

We searched the  Six degree Field Galaxy Survey
(6dFGS-DR3) database \citep{Jones+04,JSRC05,Jones+08}
for nearby galaxies in the large-scale neighbourhood of 
Abell~496, in a $\pm$0.005 redshift slice around the mean value of
0.033. This slice corresponds to $\pm$1.5 times the velocity
dispersion of a massive cluster (1000 km/s), or expressed in terms of
physical distance to a slice of 45~Mpc at the redshift of
Abell~496. On the plane of the sky, we limited our search to a box of
17$\times$17 deg$^2$ (40 Mpc at the cluster redshift).  This size is
typical of the largest cosmic bubbles \citep{HV04}.

Fig.~\ref{fig:lss}  shows that the 
6dFGS galaxies in the neighbourhood of
Abell~496 are more concentrated along a strip in the South-East/North-West
direction. This is confirmed by the distribution of groups and clusters
  that we found in NED (which is much
less homogeneous) using the same search criteria. 
Note that the 6dFGS-DR3 is complete to $K_s \leq 12.65$ \citep{Jones+08}
down to galactic latitude $|b| > 10^\circ$, and in our zone, we have $b <
-24^\circ$.   
The galaxies displayed in Fig.~\ref{fig:lss} are clearly distributed along a
large-scale filament, which appears to be at least 30 Mpc long.

We tested the prominence of this filament in the following manner. We
  searched 
for the rectangle of length $15^\circ$ (36 Mpc at the distance of Abell~496)
  and width $2^\circ$ 
encompassing the largest amount of points, imposing that Abell~496 lies along
the long axis of the rectangle, within $4^\circ$ from the closest edge along
  that axis.
We found that the furthest edge along the long axis
is at position angle (PA) $137^\circ$ 
anti-clockwise from North.
We then built 1000 random samples of as many (487) galaxies in the frame of
Fig.~\ref{fig:lss} as observed, and checked for the  most populated 
rectangle, defined as
above, covering 360 PAs in steps of $1^\circ$. While the filament in the
  observed data set has 221 galaxies within the rectangle, none of the 1000
  random datasets 
ever reached more than 81 galaxies. 
Therefore, \emph{the filament at position angle $137^\circ$ is highly
  significant.} 
This filament should constitute a
preferential avenue for infalling material into Abell~496, as well as
backsplashing material from the cluster.
However, this filament does not fully explain the excess of galaxies in the
  Southern region of Abell~496 since it is inclined relatively to the
  North-South direction.

\section{Discussion and conclusions}

\subsection{Comparison with previous analyses of Abell~496}

The LF of Abell~496 was previously measured by \cite{MCMdG98}, who
  analyzed the cluster in 4 small fields, one including the cluster centre,
  and by
  \cite{DAL02}, who measured th LF in a $42'\times28'$ field in the $I$ band.
The faint-end 
slopes of $-1.69\pm0.04$ in $r$ and $-1.49\pm0.04$ in $i$ found by
\citeauthor{MCMdG98} are consistent with our slopes of 
$-1.62\pm0.05$ and $-1.52\pm0.05$ in these bands. Moreover, 
their faint-end
slope of
$-1.34\pm0.04$ in $g$ is probably consistent with our slope ($-1.53\pm0.08$),
given that their estimated error neglects cosmic variance, which, as we show
in Fig.~\ref{fig:allsig}, is many times greater than the Poisson variance.
\cite{DAL02} find a slope of $-1.79\pm0.01$ in $I$, whereas we find here a
slope of $-1.52\pm0.05$. Given our analysis of cosmic variance, we estimate
the error from cosmic variance on the slope of \cite{DAL02} to be roughly
0.08. If this error estimate is correct, then the difference is faint-end
slopes would be statistically significant. This difference may be caused by
the different regions probed in the two studies and the different reference
fields used.

\subsection{Comparison of radial trends with other clusters}

Our $g'$, $r'$ and $i'$ LFs exhibit slopes that increase
significantly 
from the centre outwards
(Table~\ref{tab:LF}). 
This agrees with the trend for flatter LFs around cluster cDs found by
\cite{Lobo+97}, with the steeper slope in the outer envelope of Coma found
by \cite{BHvDvdH02}, and with the greater dwarf-to-giant ratio found by
\cite{DCP98}. Note, however, that all three trends are either qualitative or of
marginal statistical significance.
Although our radial trend is opposite to that found by
\cite{DePropris+03} in 
2dFGRS clusters, their shallower slope in the cluster envelopes is not
statistically significant. 

\subsection{A ubiquitous dwarf population?}
\label{ubiq}
Whereas our faint-end slope for the core of Abell 496 is fairly
shallow ($\alpha \simeq -1.4$), 
it is still steeper than most estimates of the field LF: 
\cite{Blanton+03} find $\alpha_r=-1.05\pm0.01$ for the SDSS galaxies, while
\cite{JPCS06} find $\alpha$ between $-1.10\pm0.04$ ($J$-band) and $-1.21\pm0.04$
($r$-band) for 6dFGS galaxies. However, \cite{Blanton+05} estimate that a
careful inclusion of SDSS low surface brightness galaxies yields $\alpha_r <
-1.3$ and perhaps as steep as $-1.5$.
But
several authors found very steep
slopes ($\alpha$ as steep as $-2.2$) for faint galaxy populations in
clusters (see Table~\ref{alphalit}), suggesting an important
population of dwarf galaxies in clusters, which is not seen in the
field LFs.  

It is difficult to understand how dwarf galaxies could survive
better in the hostile cluster environment than in the field.
Direct galaxy mergers should have little effect on dwarfs, whose
  cross-sections are small. Mergers of dwarfs after orbital decay by
  dynamical friction into the central cD cannot be an explanation: 
the decay
  times are expected to 
  be proportional to galaxy mass, so the giant galaxies can disappear
into the central cD. However,
the orbital decay times for the
  low mass galaxies should be long enough that the faint-end mass function
  in clusters would be the same as the field, not steeper.
Moreover, there is no observational evidence for luminosity segregation of
faint galaxies in clusters (e.g., \citealp{Pracy+05}).
Finally, tidal effects from the cluster potential 
are the same, to first order, on giant and dwarf
galaxies. Given the trend that low luminosity ellipticals are less
concentrated than more luminous ones \citep{GTC01},
the latter will
survive better the strong tides near the cluster centre, which should make
the LF shallower, not steeper.
The only possible explanation for a steeper faint-end slope in clusters would
be that galaxies of moderately low luminosity are tidally fragmented by the
cluster potential or by close encounters.

Now, if the field counts in the cluster field are underestimated
(because the reference fields are slightly underdense), the resultant
net cluster counts will be highly contaminated by field counts.
Writing the field counts as $dN/dm = \hbox{dex}[\beta\,( m-m_0)]$ and
the cluster faint-end luminosity function as $\Phi(L) \propto
L^{\alpha}$, it is easy to show that if the field dominates the net
cluster counts, one will end up measuring $\alpha=(-\beta/0.4)-1$.
For example, if $\beta=3/5$ (Euclidean counts), one would measure
$\alpha=-5/2$, while if $\beta \simeq 0.36$ (as in the DFs in the
magnitude range where we are measuring the faint-end slope of the LF),
one should find $\alpha = -1.9$.

We are confident that we suffer little from a background contamination
of our LFs. Indeed, the faint-end slopes that we have computed for the
global LF (Table~\ref{tab:LF}) and for the central region of Abell~496
(Table~\ref{tab:LF}) are all considerably and significantly shallower
than $\alpha = -1.9$ (except in the less sensitive $u^*$ band).  We
also took great care not to underestimate the counts in our field
count process by the use of very homogeneous field data covering a
large enough field of view in order to treat properly the cosmic
variance (as described above).  Finally the selection in colour that we
have applied eliminates (both in the cluster fields and in the fields
used for statistical subtraction) very red objects that are very
unlikely to belong to the cluster, so the field subtraction is quite
conservative and therefore secure.

The MegaCam colour redshift diagrams derived by \cite{Ilbert+06} from
the VIMOS VLT Deep Survey (VVDS) followup of the CFHTLS DF show that
galaxy spectra are shifted to the red to such an extent that, in the
redshift range $0.4 < z < 0.6$, the bulk of field (intrinsically blue
spiral) galaxies become redder in $g'-r'$ than nearby ellipticals, and
the same happens in the range $0.5 < z < 1.2$ for $r'-i'$ colours.  We
are therefore probably contaminated by intrinsically blue (spiral)
background galaxies with colours almost as red as our Red Sequence at
$z < 0.4$ or 0.5.

Note that two of the studies concluding to steep faint-end LF slopes
\citep{dPPHM95,Milne+07} have negative LFs at some magnitudes, which
suggests inadequate background subtraction (Fig.~\ref{fig:llf496} shows
  that 4 of our 
12 non-central local LFs also
display negative LFs at some intermediate magnitudes, which makes these 
particular LFs suspicious).  Moreover, all steep
faint-end slopes were found by authors who made no colour cuts, with
the exception of \cite{PBBR06}.  These authors estimated the LFs of
clusters in the SDSS, using the statistical subtraction method, and
found $\alpha \approx -2.2$.  The range of absolute magnitudes where
\citeauthor{PBBR06} see the rise in the faint-end LF,  [$-17.7,
    -13.7$], 
corresponds to apparent magnitudes in the range $18.03 < m < 22.03$, which
matches 
our analyzed range of apparent magnitudes.
It is puzzling that we do not find 
in our deeper images the low surface
  brightness dwarf population found by \citeauthor{PBBR06} in the SDSS images.

\begin{figure}[ht]
\centering
\includegraphics[width=9cm]{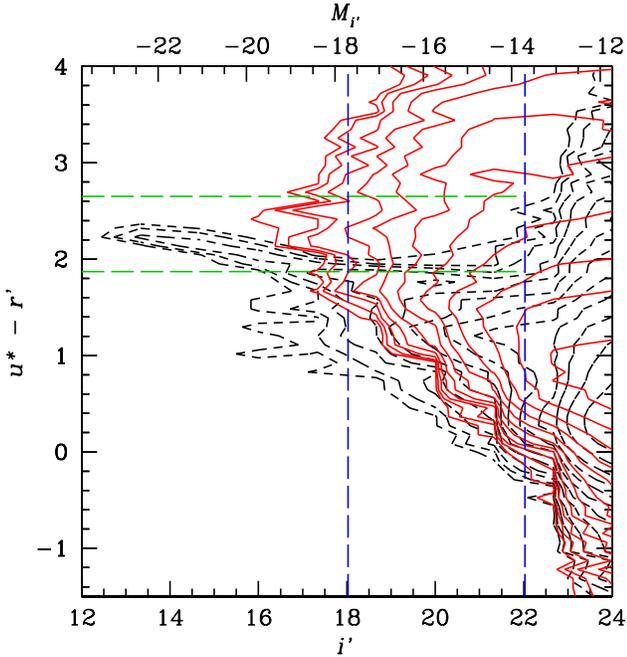}
\caption{Our $u^*-r'$ vs $i'$ colour-magnitude diagram for Abell~496.
The \emph{dashed contours} are for our selected cluster members
using equation~(\ref{coloureq}), while the \emph{solid (red)}
contours (same levels, spaced by factors of 1.6) are the galaxies
identified as field galaxies, 
because they are redder than our Red Sequence of
equation~(\ref{coloureq}).  Also shown are the limits used by Popesso
et al. (2006) to separate their red cluster galaxies and field
galaxies (\emph{upper horizontal line}) and their blue and red cluster
galaxies (\emph{lower horizontal line}), after correcting both for
$(u^*-r')_{\rm MegaCam} = (u-r)_{\rm SDSS} - 0.35$, as found using the
matches in a field of Abell~85 for which we had both SDSS 
catalogues and reduced 
MegaCam images.  The 
\emph{vertical lines} delimit the region where Popesso et al. clearly
found a steep faint-end LF slope.
\label{umrpopesso}}
\end{figure}

Fig.~\ref{umrpopesso} displays the $u^*-r'$ vs $i'$ colour-magnitude
diagram of the Abell~496 field.  
The Red Sequence is clearly visible at bright magnitudes and
its red edge is still sharp at magnitudes $18 < i' < 22$.  The
galaxies called red cluster members by \citeauthor{PBBR06} and fainter
than $M_i' = -19$ are almost all field galaxies according to our
colour-magnitude diagram.  Similarly, at $M_i' = -16.5$, our Red
Sequence (or its extrapolation) becomes bluer than the blue-red galaxy
cut of \citeauthor{PBBR06}. 
This suggests that an increasing fraction of both
\citeauthor{PBBR06}'s red and blue galaxies are in fact field
galaxies.

{}From inspection of the $u^*-g'$ and $g'-r'$ 
colour-redshift diagrams of
\cite{Ilbert+06},
one infers that the $u^*-r'$ colour cut used by \citeauthor{PBBR06} 
should vary negligibly with redshift.
This explains why
the galaxies we assign to the field (solid red contours in
Fig.~\ref{umrpopesso}) --- because they are redder than our sharp
$g'-i'$ vs $i'$ Red Sequence --- pollute the $u^*-r'$ vs $i'$
colour-magnitude diagram.  
Therefore, \emph{colour
cuts based upon $u^*-r'$ are not efficient in
  rejecting background galaxies} (with $z \la 0.4$).

Nevertheless, while LF analyses based upon the statistical subtraction
  method with
inappropriate or inexistent colour cuts 
should lead to noisier LFs, it is not clear why they should be biased to
  steeper slopes, unless the noise is such that the faint-end of the
LF fluctuates from positive to negative (which is not the case for the global
  LF of \citeauthor{PBBR06}).
Although simulations by \cite{VML01} showed that clusters selected in 2D will
  have faint-end slopes much steeper than what they put in their simulations
  ($-1.4$ instead of $-1$ in roughly the same absolute magnitude range as
  where \citeauthor{PBBR06} found their slope of $-2.3$), 
\citeauthor{VML01} have also shown that LFs
  measured in clusters
  selected in 3D, such as all the clusters reported in this paper (including
  the X-ray selected clusters studied by \citeauthor{PBBR06}), show no bias
  in the faint-end slope. A confirmation of this result is that X-ray
  selected clusters show shallower faint-end slopes than do non X-ray clusters
  \citep{VMML04}, 
  which are expected to be more prone to projection effects.

A possible cause for the discrepancy between different analyses could be the
uncertainties from star/galaxy separation. This is a serious issue for our
analysis, because Abell~496 lies at lower absolute galactic latitude ($b
= -36^\circ$) than
our reference DF fields ($b=-58^\circ, 42^\circ, 60^\circ$ and
$-53^\circ$), so that Abell~496 is more contaminated by stars.
Indeed,  our $i'$-band LF has a considerably shallower 
faint-end slope of
$-1.27\pm0.04$ instead of $-1.52\pm0.05$ (Table~\ref{tab:LF}). 
Moreover, in general,
the errors on star/galaxy
separation could go either way, leading to either an underestimation or
overestimation of the 
faint-end LF slope.

A possible explanation for the wide range of faint-end slopes (see
Table~\ref{alphalit}) is cosmic variance: some clusters may exhibit steeper
slopes than others. However, inspecting the slopes found in the literature
(Table~\ref{alphalit}), one notices a wide range of slopes for the same
cluster analyzed in similar wavebands, magnitude ranges and fields of view
(contrast 
the slopes in similar regions of Coma 
of $-1.7$, $-1.8$ of \citealp{Lobo+97} and \citealp{Trentham98a}
with the slopes shallower or equal to $-1.4$ of \citealp{AC02}).
Moreover, the two studies that have enough clusters to `beat' cosmic variance
have inconsistent 
slopes ($-2.2$ \citealp{PBBR06} and $-1.4$ \citealp{Trentham98b}) in the
magnitude range $-18$ to $-14$.

\subsection{Conclusions}

We have analyzed the galaxy LFs in the relaxed
cluster of galaxies Abell~496.  We have shown that the LFs are not only
well-defined   
in the central region (with a faint-end slope $\alpha =
-1.4\pm0.1$), but also in the South. A
concentration of clusters is indeed observed towards the South-East and
along a filament extending South-East to North-West
(Fig.~\ref{fig:lss}), suggesting the existence of a cosmological
filament linking Abell~496 with various poor clusters and groups.
However, such a filament cannot be very dense since no X-ray emission
is detected in this direction, contrary to what is observed in
Abell~85 \citep{Durret+03}.

We 
discuss the disagreement of our fairly shallow faint-end slope for
Abell~496 with that found by other authors in this and other clusters.
Although it is clear that a careful estimate of the reference field galaxy
counts and their cosmic variance are crucial, there may also be a
cosmic variance in the faint-end slopes of cluster LFs.
We find that uncertainties in the star/galaxy separation can be responsible
for some (but probably not all) of the scatter in the faint-end LF slopes given
in the literature. We highlight the removal of galaxies redder than the Red
Sequence as a means to reduce the noise in the LFs, but it is not clear if 
the lack of adequate colour cuts causes a bias towards steeper slopes.
The advent of very deep spectroscopy in cluster fields should rapidly settle
the issue of the faint-end slope of the cluster LF.

\begin{acknowledgements}
The authors thank the referee for his detailed and constructive
  comments. 
The authors would like to thank Elisabete Da~Cunha, Andrea Biviano, 
Vincent Le~Brun and Didier Pelat for useful discussions, Paola Popesso for 
useful comments and Andrea Biviano for a critical reading of the manuscript.
The authors are also grateful to the CFHT and TERAPIX teams for their help,
in particular to Emmanuel Bertin and Henry McCracken for discussions, and to 
the French PNG, CNRS for financial support.
They also thank Matthew Colless for permission to use 6dFGS-DR3, in advance
  of publication.
This research has made use of the NASA/IPAC Extragalactic Database (NED)
  which is operated by the Jet Propulsion Laboratory, California Institute of
  Technology, under contract with the National Aeronautics and Space
  Administration. 

\end{acknowledgements}


\appendix
\begin{table*}[ht]
\begin{center}
\caption{Comparison of deep photometrically-estimated
cluster galaxy luminosity functions\label{alphalit}}
\begin{tabular}{lcrll}
\hline
\hline
Cluster(s) & $r_{\rm max}/r_{100}$ & \multicolumn{1}{c}{abs. mag range}  & \multicolumn{1}{c}{$\alpha$} & \multicolumn{1}{c}{Reference} \\
\hline
\object{Local Group} & 3.6 & $M_V < -9$ & --1.1 & \cite{PvdB99} \\
\object{Virgo} & 1.1 & $M_B < -13.1$ & --1.25 & \cite{SBT85} \\
Virgo & 1.1 & $M_B < -11.1$ & --1.3 &
\cite{SBT85} \\ 
Virgo & $0.24\!\times\! 0.24$ & $-15.6 < M_R < -11.1$ & --2.26 &
\cite{PPSJ98} \\
Virgo & $2\!\times\!0.4\!\times\!1.2$ & $-18 < M_B < -11$ &
--1.35 & \cite{TH02} \\ 
Virgo & $2\!\times\!0.4\!\times\!1.2$ & $-17 < M_B < -14$ &
--1.7 & \cite{TH02} \\ 
\object{Coma} & $0.08\!\times\! 0.08$ & $M_R < -11.6$ & --1.42 &
\cite{Bernstein+95} \\
Coma & $0.08\!\times\! 0.08$ & $-11.6 < M_R < -9.4$ & --2.0 &
\cite{Bernstein+95} \\
Coma & $0.58\!\times\! 0.24$ & $M_V < -15.5$ & --1.8 &
\cite{Lobo+97} \\ 
Coma & $0.3\!\times\!0.3$& $-15.6 < M_B < -10.6 $ & --1.7&
\cite{Trentham98a} \\
Coma & $0.3\!\times\!0.3$& $-17.6 < M_R < -11.6 $ & --1.7&
\cite{Trentham98a} \\
Coma & 0.8 & $M_U < -13.4$ & --1.32 & \cite{BHvDvdH02} \\
Coma & 0.8 & $M_B < -13.4$ & --1.37 & \cite{BHvDvdH02} \\
Coma & 0.8 & $M_r < -13.4$ & --1.16 & \cite{BHvDvdH02} \\
Coma & $0.28\!\times\! 0.43$ & $M_{B,V,R} < -12.8$ & --1.25 & \cite{AC02} \\
Coma & $0.28\!\times\! 0.43$ & $M_V < -11.3$ &--1.4 & \cite{AC02} \\
Coma & $0.28\!\times\! 0.43$ & $M_R < -11.8$ &--1.4 & \cite{AC02} \\
Coma & $0.6\times0.6$ & $-19.1 < M_R < -14.6$ & --1.55 &
\cite{IglesiasParamo+03} \\  
Coma & 0.01 & $M_R < -9.1$ & --2.29 & \cite{Milne+07} \\
Coma & 0.01 & $M_R < -11.3$ & --1.9 & \cite{Adami+07} \\
Coma (North)
& $0.28\!\times\! 0.22$ & $M_B < -10.5$ &--1.48 & \cite{Adami+07} \\
Coma (North) 
& $0.28\!\times\! 0.22$ & $M_R < -11.3$ &--1.74 & \cite{Adami+07} \\
Coma (South)& $0.28\!\times\! 0.22$ & $M_B < -10.5$ &--1.32 &
\cite{Adami+07} \\ 
Coma (South) & $0.28\!\times\! 0.22$ & $M_R < -11.3$ &--1.28 &
\cite{Adami+07} \\ 
\object{Abell 426} & $0.1\times0.1$ & $-19.4 < M_I <
  -13.4$ & --1.56 & \cite{dPP98} \\
\object{Abell 496} & $4\!\times\!0.19\!\times\!0.19$ & $M_g < -13.2$ & --1.34 &
\cite{MCMdG98} \\
Abell 496 & $4\!\times\!0.19\!\times\!0.19$ & $M_r < -13.2$
& --1.69 & \cite{MCMdG98} \\
Abell 496 & $4\!\times\!0.19\!\times\!0.19$ & $M_i < -13.2$
& --1.49 & \cite{MCMdG98} \\
Abell 496 & $0.9\!\times\!0.6$ & $M_I^{\rm AB} < -13.7$ & --1.79 & \cite{DAL02} \\
\object{Abell 539} & $0.24\times0.24$ & $-18.5 < M_I <
  -14.0$ & --1.42 & \cite{dPP98} \\
\object{Abell 1185} & 0.9 & $M_B < -12.4$ & --1.25 & \cite{ACPM06} \\
Abell 1185 & 0.9 & $M_V < -13.2$ & --1.28 & \cite{ACPM06} \\
Abell 1185 & 0.9 & $M_R < -13.7$ & --1.28 & \cite{ACPM06} \\
\object{Abell 1367} & $0.7\times0.7$ & $M_R < -14.3$ & --1.07 &
\cite{IglesiasParamo+03} \\ 
\object{Abell 2199} & $0.04\times0.04$ & $M_B < -10.5$ & --2.16 &
\cite{dPPHM95} \\ 
3 clusters & $0.05\times0.05$ & $M_I < -13.0$ & --2.28 &  \cite{dPPHM95} \\
9 clusters & $0.25\times0.25$ & $-19 < M_B < -14$ & --1.4 & \cite{Trentham98b}
\\
9 clusters & $0.25\times0.25$& $-14 < M_B < -11$ & --1.8 & \cite{Trentham98b} \\
\object{Ursa Major} & 6.8 & $-17 < M_R < -11$ & --1.1 & \cite{TTV01} \\
69 RASS/SDSS clusters & 0.7 & $M_g < -13.7$ & --1.98 &
\cite{PBBR06} \\
69 RASS/SDSS clusters & 0.7 & $M_r < -13.7$ & --2.19 &
\cite{PBBR06} \\
69 RASS/SDSS clusters & 0.7 & $M_i < -13.7$ & --2.26 &
\cite{PBBR06} \\
69 RASS/SDSS clusters & 0.7 & $M_z < -13.7$ & --2.25 &
\cite{PBBR06} \\
\hline
\end{tabular}
\end{center}
Notes: The method used is field subtraction, except for the Local Group study
of \cite{PvdB99} and the first line of the Virgo analysis by \cite*{SBT85},
which are based upon raw counts, while the second line of the Virgo analysis
of \citeauthor{SBT85} adds a statistical background correction.
The virial radii are from \cite{MSSS04} (Virgo), \cite{LM03} (Coma), 
this paper (Abell~496),
\cite{MVFS99} (Abell~2199), and otherwise adapted from the velocity
dispersions measured 
by \cite{Fadda+96}, when available (only for 5 out of the 9 clusters of
  \citealp{Trentham98b} and 3 of 4 for \citealp{dPPHM95}). 
Magnitude ranges are converted to $H_0 = 72 \,\rm km \,s^{-1} \,
Mpc^{-1}$. Typical errors on $\alpha$ are in the range 0.1 to 0.2.
\end{table*}

\begin{table*}
\centering
\caption{Luminosity functions for Abell 496 in the whole field
(Fig.~7) and in the three main areas (Fig.~9).}
\begin{tabular}{r*4{r@{$\ \pm\ $}l}}\hline\hline
mag bin
& \multicolumn{2}{c}{$u^*$} & \multicolumn{2}{c}{$g'$} & 
  \multicolumn{2}{c}{$r'$}  & \multicolumn{2}{c}{$i'$} \\ \hline
& \multicolumn{8}{c}{All} \\
 13.5 &   0 &   0 &   1 &   0 &   2 &   0 &   7 &   1 \\     
   14 &   0 &   0 &   5 &   1 &  11 &   1 &  24 &   1 \\      
 14.5 &   0 &   0 &  15 &   1 &  22 &   1 &  24 &   1 \\      
   15 &   0 &   0 &  23 &   1 &  23 &   1 &  10 &   1 \\     
 15.5 &   2 &   1 &  24 &   1 &  18 &   1 &  24 &   1 \\     
   16 &  12 &   1 &  24 &   2 &  28 &   2 &  39 &   2 \\     
 16.5 &  20 &   2 &  39 &   3 &  44 &   3 &  28 &   3 \\     
   17 &  30 &   2 &  49 &   3 &  15 &   3 &  24 &   4 \\     
 17.5 &  24 &   2 &  35 &   4 &  27 &   4 &  42 &   5 \\     
   18 &  27 &   3 &  29 &   6 &  41 &   6 &  45 &   7 \\     
 18.5 &  46 &   3 &  50 &   9 &  69 &   9 &  58 &  11 \\    
   19 &  32 &   5 &  61 &  13 &  51 &  14 &  72 &  17 \\   
 19.5 &  38 &   9 &  63 &  20 &  91 &  24 & 106 &  28 \\    
   20 &  80 &  25 & 103 &  26 & 104 &  28 & 101 &  33 \\    
 20.5 &  76 &  36 & 100 &  31 & 118 &  46 & 119 &  47 \\    
   21 & 125 &  65 & 161 &  53 & 174 &  63 & 152 &  72 \\   
 21.5 & 156 & 104 & 188 &  73 & 233 &  98 & 243 & 117 \\   
   22 & 239 & 172 & 251 & 132 & 258 & 166 & 321 & 177 \\   
 22.5 & 328 & 308 & 390 & 228 & 536 & 257 & 583 & 304 \\  \hline
& \multicolumn{8}{c}{Center} \\ 
17 &   85 &    4 &  165 &   7 &  126 &   9 &  110 &  11 \\
18 &  112 &    6 &  131 &  13 &  177 &  16 &  195 &  22 \\
19 &  153 &   10 &  229 &  29 &  218 &  35 &  283 &  44 \\
20 &  232 &   42 &  336 &  61 &  378 &  67 &  306 &  82 \\
21 &  382 &  132 &  372 & 102 &  520 & 137 &  461 & 154 \\
22 &  594 &  356 &  637 & 251 &  733 & 311 &  734 & 355 \\
23 & 1895 & 1264 & 2256 & 813 & 1922 & 910 & 1387 & 949 \\ \hline
& \multicolumn{8}{c}{South} \\ 
17 &   39 &    2 &   88 &    7 &   32 &   6 &   25 &    8 \\
18 &   36 &    6 &   20 &   15 &   47 &  15 &   49 &   18 \\
19 &   33 &   13 &   94 &   32 &  119 &  34 &  150 &   40 \\
20 &  157 &   44 &  201 &   64 &  225 &  75 &  221 &   76 \\
21 &  252 &  126 &  347 &  109 &  417 & 150 &  535 &  174 \\
22 &  578 &  358 &  757 &  351 &  917 & 384 &  976 &  431 \\
23 & 2605 & 1337 & 3275 & 1009 & 2182 & 991 & 2444 & 1065 \\ \hline
& \multicolumn{8}{c}{East-North-West} \\ 
17 &   54 &    2 &   64 &   6 &  22 &   7 &  22 &    8 \\
18 &   33 &    3 &   46 &  13 &  64 &  16 &  70 &   17 \\
19 &   44 &    8 &   67 &  30 &  76 &  36 &  70 &   38 \\
20 &   76 &   47 &  106 &  56 & 110 &  69 & 122 &   79 \\
21 &  121 &  126 &  157 & 128 & 202 & 145 & 204 &  151 \\
22 &  314 &  347 &  313 & 316 & 450 & 363 & 406 &  378 \\
23 & 1276 & 1262 & 1308 & 991 & 992 & 991 & 720 & 1038 \\ \hline\hline
\end{tabular}
\end{table*}
\end{document}